# Towards a Precise Semantics for Object-Oriented Modeling Techniques*

Ruth Breu, Radu Grosu, Franz Huber,
Bernhard Rumpe, Wolfgang Schwerin

Institut für Informatik
Technische Universität München
email: {breur,grosu,huberf,rumpe,schwerin}@informatik.tu-muenchen.de

**Abstract** In this paper, we demonstrate how a precise semantics of object-oriented modeling techniques can be achieved, and what the possible benefits are. We outline the main modeling techniques used in the SYSLAB project, sketch, how a precise semantics can be given, and how this semantics can be used during the development process.

## 1 Introduction

The development of complex software systems is a subject of great technical, economic and scientific importance. A software development method can be defined as a unified approach incorporating multiple description techniques, characterising a system from several points of view. Most of these description techniques currently used, however, lack a formal semantics. While recent research works on *formal methods* aim at the formal foundation of separate description techniques, less emphasis is put on the formal integration of the multiplicity of description techniques used in a single method. Yet, integrated description techniques are the basis for a systematic design and for vast tool support during the development process.

Besides the use of description techniques in specific methods, one has to consider them in a more general scientific context. Their importance for modeling software systems might turn out to be comparable to the importance of mathematical techniques, invented in the second half of the 19'th century to model physical processes. Therefore, a scientific foundation of description techniques seems to be of great significance.

It is the aim of the SYSLAB project to develop a mathematically founded modeling technique for distributed, object-oriented systems, based on UML [BRJ96] description techniques. The modeling technique will offer a systematic set of steps for enhancing, refining, and transforming *documents* of the description techniques used in SysLab. It supports the systems development process from analysis to implementation.

---

* This paper originates from the SYSLAB project, which is supported by the DFG under the Leibnizprogramme and by Siemens-Nixdorf.

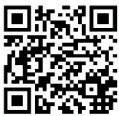

## 2 The SysLab Description Techniques

### 2.1 Modeling Method

A modeling method roughly defines the process of software development. It turns out that the description techniques used and their usage order are rather orthogonal. It therefore makes sense to develop the description techniques and their precise semantics independently of the modeling method, as, e.g., done in UML 1.0. However, the semantics has a severe impact on the possible transformation steps for documents. These transformation steps are the connection between the description techniques and the method. A method can be seen as a set of guidelines and heuristics that tell the developer when and why to use a sequence of transformations. The method tells what the prerequisites are, what the benefits are, and what pitfalls should be avoided (quite similar to design patterns [GHJV94]).

Description techniques used to define different views of a system, play a central role within a modeling method. Documents describing a system using these techniques are used and transformed until the whole system is described by a set of executable documents. Basically, we use the following description techniques originating from UML, but adapted and specialized to allow the definition of a precise semantics:

- Informal Text and Diagrams (ITD)
- Message Sequence Charts (MSC)
- State Transition Diagrams (STD)
- Object Model (OM)
- Specification Language (SL)
- Programming Language (PL)

Documents of these kinds are provided with a semantics based on a *mathematical system model* (MSM). Through this semantic foundation, we not only get a precise semantics for documents, but also an integrated one, which allows us to define transformations between documents as well as rigorous context conditions within and between different description techniques.

A transformation step takes a finite set of documents (often one) and produces new documents. The set of possible transformations is to be chosen carefully, to ensure systematic and correct manipulation of documents. Then it is, e.g., possible to inherit the STD-based behaviour description of a class to its subclasses using a refinement calculus, as, e.g., given in [Rum96, RK96], which is similar to refinement calculi, like e. g. the work of C. Morgan [Mor90].

The development of a system is captured in a *development graph*, which contains documents as nodes and dependencies between them as directed arcs. Each document has a state which, e.g., captures whether a document is still necessary or already redundant, because its successor documents contain all information of the document. Such information for documents is necessary, on the one hand, to trace requirements and design decisions through the development process, and, on the other hand, to allow requirement changes in a systematic way.

## 2.2 Description Techniques

For software engineers it is extremely important to describe complex structural and dynamic dependencies in a clear, structured and systematic way. Therefore, several description techniques, providing different *views* as well as different *abstraction levels*, are used.
Based on existing object-oriented modeling techniques like UML or OMT [BRJ96, RBP$^+$91] we use the following techniques as core of the SysLab-method:

*Informal Text and Diagrams (ITD)* comprises any kind of text, diagrams, tables and graphics. Whenever desired or necessary, ITD can be used, thus allowing scalability of formal techniques. It is escpecially useful to capture requirements in early phases, comments and annotations not yet fully explored, and to store reasons for design decisions. Despite its informal character, ITD can be used in a systematic way, e. g. to extract of classes and attributes from requirements descriptions. We also attach a state to informal documents, capturing e.g. the validation or redundancy state of a document.

*Message Sequence Charts (MSC)* describe separate flows of communication or subsets of communication flows in a system. Emphasis is put on communication between separate parts (objects or groups of objects) of a system. Constituting a high level of abstraction, MSCs are well suited to capture a system's requirements. Moreover, MSCs can be used for and generated by simulation respectively. Our MSC variant is based on the message-oriented model and allows us to define different layers of abstraction, repetition, choice and hierarchy of MSCs.
One of the main and still not completely explored problems is the semantics of an MSC in the presence of underspecification and nondeterminism. It seems, that some kind of completeness assumption could be necessary to allow a set of MSCs to be given a semantics. Furthermore, a starting part (usually the first message) will be considered as a starting trigger.

*State Transition Diagrams (STD)* describe the lifecycle of objects. In STDs, descriptions of state and behaviour are combined. Different levels of abstraction allow both the specification of an object's interface as well as the specification of methods. Refinement techniques enable not only inheritance of behaviour but also stepwise refinement of abstract STDs, resulting in an implementation.
To describe a detailed behaviour of transitions, it is necessary to use a specification language that relates input and source state with output and destination state. This specification language (SL) is characterised below.

*Object Model (OM)* describes the static structure of a system. The OM encompasses the description of classes and of relationships between classes: association, aggregation, and generalization. It includes the signature of objects, given by their operations and attributes.
To describe structural invariants that have to be maintained, we use the same specification language as for transitions in STDs.

*Specification Language (SL)* is an axiomatic specification language based on predicate logic, resembling Spectrum [BFG$^+$93]. SL allows declarative definition of properties. Particularly, SL is used for the definition of pre- and postconditions of transitions and for the definition of state invariants not only in single objects but also between several objects in the OM. In order to enable automatic testing of verification conditions, SL is also oriented towards functional programming, resembling Gofer [Jon93] in this concern. As an effect, the step from high-level descriptions towards executable code is facilitated, which again makes prototyping easier.

*Programming Language (PL)* is an executable implementation language. System descriptions formulated in an executable language are the target of any software development process. Therefore the integration of PL in our method is a must. Designing PL as a subset of the object-oriented language Java [Fla96] seems to be reasonable. Besides others, Java has the advantage of being architecture-independent. In order to fully integrate PL into the development process, assigning PL a formal semantics is necessary.
For each description technique, except informal documents (ITD), a formal *abstract syntax*, a *concrete diagrammatic or textual representation*, and a complete set of *context conditions* for the correctness of documents will be supplied. Furthermore, a *formal semantics* based on the MSM will be given.

## 3  Mathematical System Model (MSM)

### 3.1  Informal Description of MSM

The mathematical system model serves as a basis for the creation of the semantics of the description-techniques. The MSM describes the universe of systems $\mathbb{SM}$ that can be specified by the SYSLAB-method. The MSM is formalized using mathematical techniques [RKB95, KRB96]. However, for an understanding of the SYSLAB method it is not necessary to know the formalization of the MSM. For this reason, we only roughly sketch the MSM below.
A system consists of a dynamically changing set of *objects*, each with its own *identity*. The *objects* are grouped by a finite set of *classes*. A *state* is assigned to each object. Both the object's *attributes* as well as the set and states of its active operations determine the object's state. A *signature* describes the set of incoming and outgoing *messages*, which can be classified into *method calls* and *return messages*.

## 3.2 Formalisation and Usage of MSM

Let $\mathbb{SM}$ be the set of systems that we are interested in. Let us assume that we have formalised the syntax for the description techniques, resulting in a set of context correct documents $\mathbb{DOC}$. The semantics of one document $d \in \mathbb{DOC}$ is given by a set of systems that obey the restrictions of this documents. Formally, we define the semantics function as:

$$[[.]] : \mathbb{DOC} \to \mathbb{P}(\mathbb{SM})$$

If, for example, $d$ is an object model, each class mentioned in $d$ has to exist in each system $s \in [[d]]$. Classes not mentioned in $d$ may exist, but need not. A canonical minimal system may be implemented containing only mentioned classes, but adding new classes is a perfect refinement. This "loose", set based semantics [BBB+85] for documents allows a very simple and powerful extension of the semantics function to sets of documents $D \subseteq \mathbb{DOC}$:

$$[[D]] \stackrel{def}{=} \bigcap_{d \in D} [[d]]$$

This definition captures the idea that adding documents, and thus refining the existing information about the system in development, rules out more and more systems, until only the system to be implemented remains as semantics.

We now can define the notion of redundancy. A document $d$ is redundant with respect to another document $d'$, if the semantics of the latter is a precision of the former: $[[d']] \subseteq [[d]]$. Any redundant document does not need to be considered in the development any longer, as the semantics of the complete set of documents is the same as of the non redundant subset, here: $[[d, d']] = [[d']]$.

However, a document being redundant in this formal sense can still be important for documentation reasons. For example, the abstract and therefore redundant version of a document may omit details which are irrelevant for human understanding. We can also define the notion of refinement $\models$. Document $d'$ refines $d$, is defined by:

$$d \models d' \stackrel{def}{=} [[d']] \subseteq [[d]]$$

This definition immediately shows, that the refined document $d$ becomes redundant, and only the refinement $d'$ has to be considered furthermore.

This notion of semantics allows us to classify different kinds of transformations of documents. We can for example distinguish between transformations that add information and are therefore true refinements or semantics preserving transformations.

On the one hand, these transformations must grant as much freedom as possible to the developer. On the other hand, a systematic development of correct systems has to be ensured.

## 4   Conclusion

In this paper, a coherent set of description techniques based on UML and used in the SysLab project has been presented. Documents, created using these description techniques specify a set of systems in a loose manner. The development of a system can be understood as the repeated transformation, e.g., refinement, of documents. The introduction of a mathematical system model assigns not only an *integrated* formal semantics to the set of description techniques, but also to the set of transformations.